\documentclass[]{raa}            
\usepackage{graphicx,times}
\usepackage{natbib}

\providecommand{\bm}[1]{\mbox{\boldmath$#1$\unboldmath}}

\begin{document}

   \title{No Enough Evidence In Support Of Correlation Between Gamma-ray Bursts And Foreground Galaxiy Clusters In the Swift Era 
}

 \volnopage{ {\bf 2009} Vol.\ {\bf 9} No. {\bf XX}, 000--000}
   \setcounter{page}{1}

   \author{Jing Wang
      \inst{1}
   \and Jian-yan Wei
      \inst{1}
   }

   \institute{National Astronomical Observatories, Chinese Academy of Sciences,
             Beijing 100012, China; {\it wj@bao.ac.cn}\\
\vs \no
   {\small Received [year] [month] [day]; accepted [year] [month] [day] }
}

\abstract{The correlation between distant Gamma-Ray Bursts (GRBs) and foreground galaxy clusters is 
re-examined by using the well localized (with an accuracy down to a few arcseconds) 
\it Swift/\rm XRT GRBs. The galaxy clusters are compiled from both X-ray selected \it ROSAT \rm 
brightest cluster sample (BCS) and BCS extension by requiring $\delta \geq0\degr$ and 
$b\geq20\degr$. The \it Swift/\rm XRT GRBs fulfilling the above selection criteria are 
cross-correlated with the clusters. Both Nearest-Neighbor Analysis and angular two-point 
cross-correlation function show that there is no enough evidence supporting the correlation 
between the GRBs and foreground clusters. We suggest that the non-correlation is probably related
to the GRB number-flux relation slope. 
\keywords{gamma rays: bursts -- Galaxies: clusters: general -- methods: statistical
}
}

   \authorrunning{J. Wang \& J. Y. Wei}            
   \titlerunning{No Evidence of GRB/Cluster Correlation}  
   \maketitle


%
%
\section{Introduction}           
\label{sect:intro}

It is now no doubt that Gamma-Ray Bursts (GRBs) take place at cosmological distance.
Thanks to the prompt localizations and deep follow-up observations,
the record of the highest redshift of GRBs has been progressively broken in past a few years,
especially after the launch of the \it Swift \rm satellite (Gehrels et al. 2004).
At present, GRB\,090423 detected by the \it Swift \rm satellite is the most
distant GRB with a redshift of $\sim8.1$ (Salvaterra et al. 2009). So far, there are 
about 50 \it Swift \rm detected GRBs with measured redshifts. A majority of these GRBs 
lie beyond $z=1$ with a redshift distribution that peaks at $z\sim1-2$.

The cosmological origin and high luminosities offer an opportunity to use GRBs as tracers
to study: star formation history of the Universe (e.g., Savaglio et al. 2009; 
Jakobsson et al. 2005); properties and evolution of intergalactic medium and high redshift galaxies 
(e.g., Prochaska et al. 2006;
Prochter et al. 2006; Vergani et al. 2009), similar as done with the high redshift quasars;
and the nearby mass distribution through the weak lensing of GRBs caused by the 
local large-scale structure (Williams \& Frey 2003 and references therein).

Because 
GRBs are point-like sources, their weak lensing could only be detected trough the angular 
correlation between sources and corresponding lenses\footnote{Another way looking for candidates of 
GRB lensing effect is based on the time delay of two bursts from the same sky region 
(e.g., Veres et al. 2009)}.  
A number of authors previously examined whether
subsets of GRBs are correlated with subsets of foreground galaxy clusters. The results obtained by
these authors are, however, contradictious. Kolatt \& Piran (1996) claimed the 136 GRBs
selected from the Burst and Transient Source Experiment (BATSE) 3B catalog are correlated with
the Abell cluster (Abell et al. 1989) within an angular separation of 4\degr at a
significance level 95\%. Marani et al. (1997) obtained a stronger correlation by using
the BATSE GRBs with more accurate positions. By contrast, Hurley et al. (1999) did not
find any evidence for the correlation between GRBs and galaxy clusters by extending the GRB
sample to the BATSE 4B/Third Interplanetary Network catalog. In addition, Williams \& Frey (2003)
reported an anti-correlation between the BATSE GRBs and Abell clusters.

A caveat in these previous studies is the large BATSE error box that usually ranges from a fraction of a 
degree to a few degrees.
The error box in some cases is as high as $\sim30\degr$.
The poor localization has been greatly improved after the
launch of the \it Swift \rm satellite. The spacecraft of the \it Swift \rm satellite can quickly slew
to the GRB position given by the BAT instrument within 100 seconds.
Due to the high sensitivity, the XRT onboard the \it Swift \rm satellite has the capability that
measure X-ray afterglow position with an accuracy better than 5\arcsec within
100 seconds for about 90 percent BAT triggers (Burrows et al. 2005).

Here, we re-examine the correlation between GRBs and foreground galaxy clusters
by using the \it Swift\rm/XRT sample. As mentioned before, the highly accurate position
provided by the XRT instrument allows us to regard these GRBs as point sources. The correlation
is studied by the nearest-neighbor distance method and angular two-point cross-correlation
function. Both methods indicate that there is no significant correlation between the GRBs
and foreground galaxy clusters. ....
The paper is organized as follows. \S 2 describes the sample selection. The analysis and
results are presented in \S 3. A short discussion and a conclusion are provided in \S4.


\section{Sample Selection}
\label{sect:sample}

The correlation between GRBs and foreground galaxy clusters are investigated in the current study by using
both X-ray selected \it ROSAT \rm brightest cluster sample (BCS, Ebeling et al. 1998) and BCS extension
(Ebeling et al. 2000). Both samples are selected from
the \it ROSAT  \rm All-Sky Survey (RASS) in the northern hemisphere ($\delta\geq 0\degr$) and
at high Galactic latitude ($|b|\geq20\degr$). By combining the two samples, the completeness is about 75\% down to
a total flux limit of $2.8\times10^{-12}\ \mathrm{erg\ s^{-1}\ cm^{-2}}$. There are totally 203 and
107 clusters in the BCS and BCS extension, respectively. Among the combined sample, 300 clusters
have measured redshifts less than 0.3 (see Figure 3 in Ebeling et al. 2000). The \it Swift\rm/XRT GRBs
are selected from the current \it Swift\rm/XRT catalog \footnote{The
catalog can be obtained from the web site http://heasarc.gsfc.nasa.gov/docs/swift/archive/grb\_table/.}.

To test the association of GRBs with foreground galaxy clusters,
we further select a subset of GRBs and a subset of
clusters from the selected samples given above by requiring $\delta\geq 0\degr$ and $b\geq20\degr$.
Our samples used in this study
are finally comprised of 116 \it Swift\rm/XRT detected GRBs (hereafter GRB sample for 
short) and 223 clusters. A sub-sample of 54 GRBs (hereafter GRBz sample) 
with measured redshifts (both spectroscopic and photometric redshifts) is extracted from the GRB sample. 
Figure 1 shows
the distributions of the GRBs and clusters on the sky in the equatorial coordinate. The
total sky area covered by the cluster sample and by the GRB sample is calculated to be
9684 square of degree.

Figure 2 shows the redshift distribution of the GRBs listed in the GRBz sample. Note that all the GRBs have
redshifts larger than 0.3, which greatly differs from the redshift distribution of the clusters,
except two outliers: GRB\,060502B (at $z=0.287$) and GRB\,050509B (at $z=0.225$).
The comparison of the redshift distributions between the GRBs and clusters indicates that the
relation between the GRBs and clusters, if any,
is unlikely affected by their physical association. In fact, our Nearest-Neighbor Analysis 
(NNA, see below) shows that the redshifts of the closest clusters associated with GRB\,060502B and
GRB\,050509B are 0.0473 (Zwicky\,8338) and 0.1997 (Abell\,1602), respectively.

\begin{figure}[h!!!]
\centering
\includegraphics[width=9.0cm, angle=0]{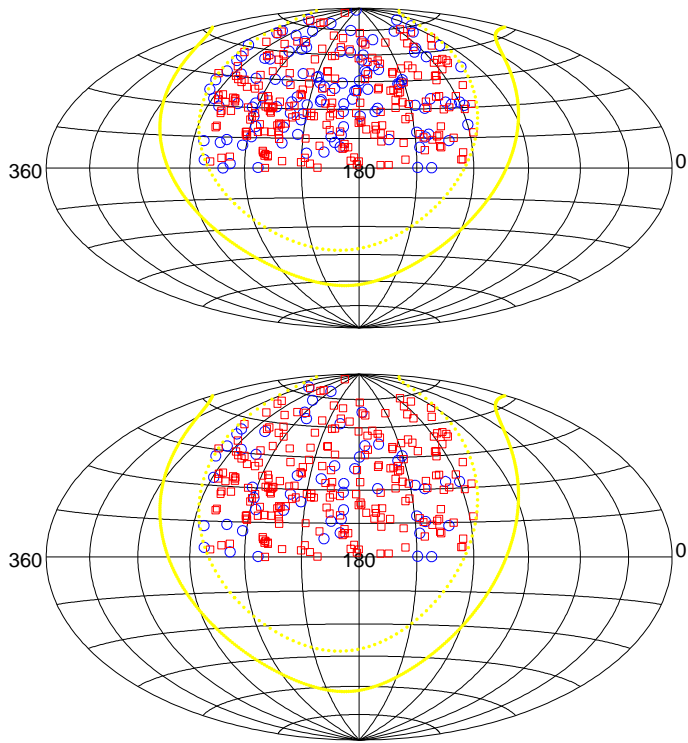}
\begin{minipage}[]{85mm}
\caption{The distributions of the samples used in this study on the sky in the equatorial
coordinate. The GRBs are marked by the blue open circles. The upper and lower panels show
GRB sample and GRBz sample, respectively. 
The solid and open red squares present
the used foreground clusters listed in the BCS and BCS extension, respectively.
The equator plane of the Galactic
coordinate and plane with Galactic latitude of $b=20\degr$ are marked by a yellow solid and a yellow dotted lines, respectively} 
\end{minipage}
\label{Fig1}
\end{figure}

\begin{figure}[h!!!]
\centering
\includegraphics[width=9.0cm, angle=0]{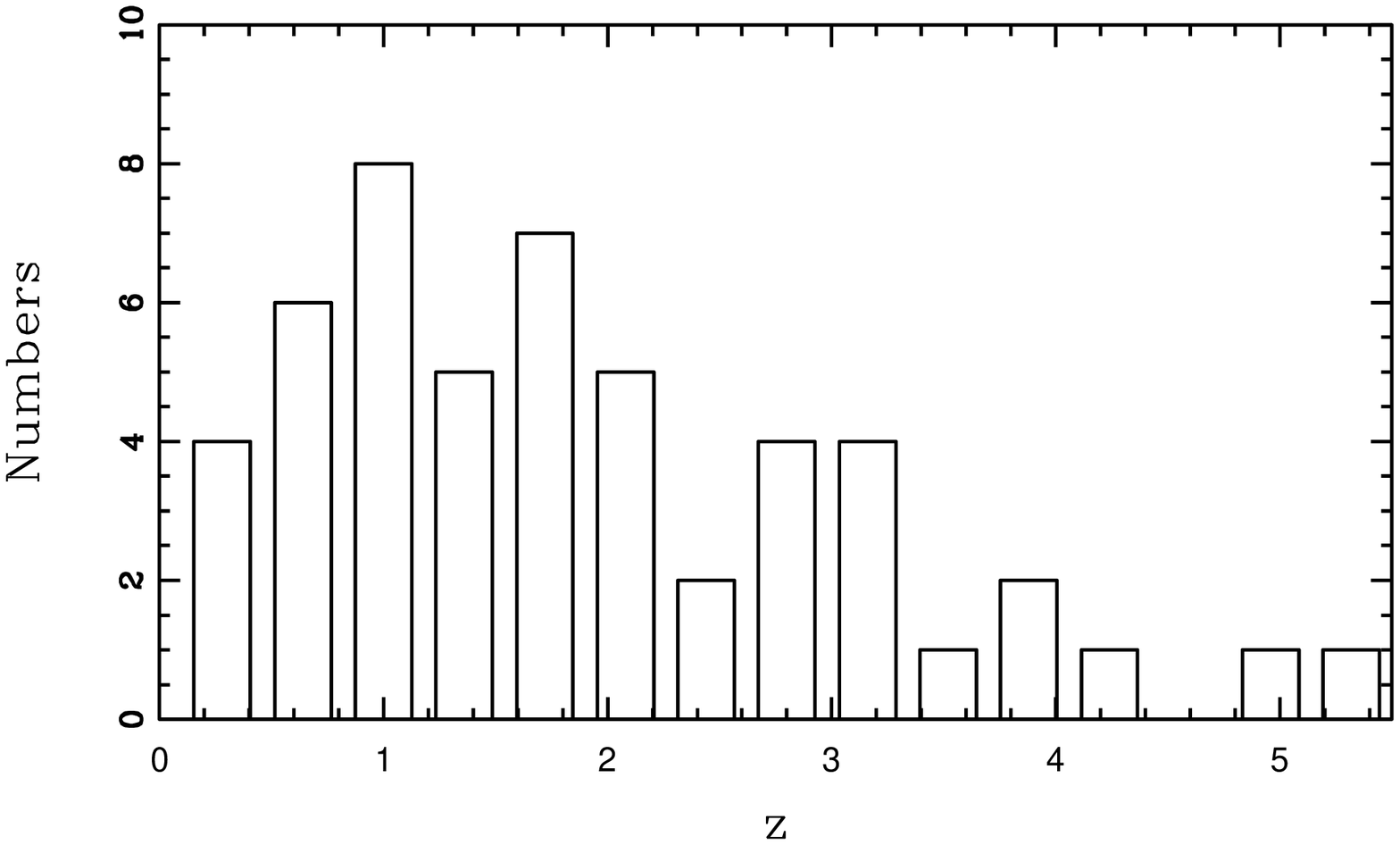}
\begin{minipage}[]{85mm}
\caption{
Redshift distribution of the 54 \it Swift/\rm XRT GRBs used in this study.
}
\end{minipage}
\label{Fig1}
\end{figure}

\section{Analysis And Results}
\label{sect:ana}

In this section, we re-examine the correlation between GRBs and foreground galaxy clusters in terms of
the samples established in the above section. Both NNA and 
angular two-point correlation function are adopted in our analysis.

\subsection{Nearest-Neighbor Analysis}

NNA is believed to be insensitive to the inhomogeneous distribution of 
samples on large scales. It has been widely used in searching for anisotropis in astronomy 
(e.g., Bachall \& Soneira 1981; Impey \& He 1986; Yang et al. 1995; Meszaros \& Stocek 2003). 
The main advantage of NNA is its simplicity that the method uses only 2$N$ angular distances (pairs).
In addition to the simplicity,   
the nearest-neighbor distance directly reflects the association of two types
of objects. The null hypotheses of the NNA is that 
both GRBs and clusters are uniformly distributed on the sky with the Poisson statistics. 
The hypotheses means that the probability of
finding the nearest neighbour in the range between $\theta$ and $\theta+d\theta$ can be described as 
following probability density function that was rigorously derived in Scott \& Tout (1989): 

\begin{equation}
P(\theta)d\theta=2\pi n\theta e^{-\pi n\theta^2}d\theta
\end{equation}
where $n$ is the surface density of the objects.

The solid line shown in each panel of Figure 3 plots the calculated empirical
distribution of the nearest-neighbor distances. The distributions are plotted by 
binning the data into an angular bin size of 1\degr.
The left column shows the distributions of 
the nearest-neighbor distances of the clusters, GRB sample, and GRBz sample. 
The right column shows the distributions of distance between the clusters and GRB sample, and 
distance between the clusters and GRBz sample.
Each empirical distribution is compared 
with a simulated distribution shown by the heavey dashed line. Each simulated distribution 
is obtained through 500 random realizations of GRB/cluster distributions on the sky, and 
binned into the same angular bin size. The every 
random catalog has not only the same number of the real data, but also the same sky coverage.
The two sets of the thin dashed lines in each panel mark the
1$\sigma$ and 2.5$\sigma$ confidence limit derived from the simulation by assuming a Guassian distribution.
The Guassian distribution is, however, no longer a good approximation when only a small number of events 
are detected. In these cases (typically, the events are less than 10), the error tables given in Gehrels (1986) 
are adopted here to provide reasonable limits.
  
Clearly, one can see that all the empirical distributions are significantly
consistent the expected distributions if the GRBs and clusters are randomly distributed,
except the case of the clusters (see the upper-left panel). 
There is a significant
clustering (with a significance level larger than 99\%) for the clusters when the angular 
distance between two clusters is less than 3\degr, 
although the clusters become to be randomly distributed on the sky on larger scales.
In fact, this significant clustering is well confirmed by our subsequent analysis basing upon 
two-point correlation function (see Figure 4).
This result is qualitatively in agreement with the previous studies as well
(e.g., Zandivarez et al. 2001; Bahcall \& West 1992; Croft et al. 1997). 

\begin{figure}[h!!!]
\centering
\includegraphics[width=9.0cm, angle=0]{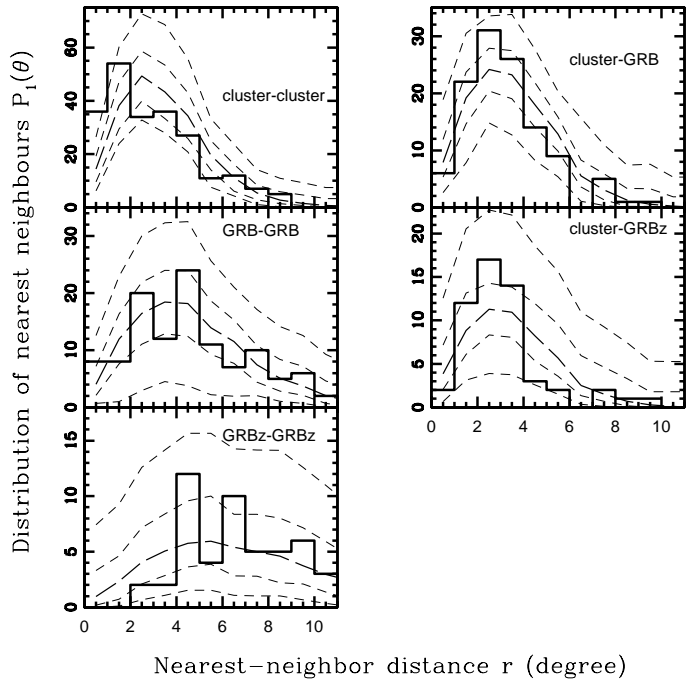}
\begin{minipage}[]{85mm}
\caption{The calculated empirical distributions of the nearest-neighbor distances are 
compared with the Monte-Carlo simulations with 500 random realizations. The left column shows the distributions of
the nearest-neighbor distances of the clusters, GRB sample, and GRBz sample.
The right column shows the distributions of distance between the clusters and GRB sample, and
distance between the clusters and GRBz sample. In each panel, the calculated empirical distribution 
of the nearest-neighbor distance is presented by the solid line, and the Monte-Carlo simulation
by the dashed lines. The two sets of the thin dashed lines in each panel mark the
1$\sigma$ and 2.5$\sigma$ confidence limit derived from the simulation.
} 
\end{minipage}
\label{Fig1}
\end{figure}


\subsection{Angular Two-point Correlation Function}

The two-point correlation function is a commonly used statistical tool for 
researching the large scale structure. The angular two-point correlation function $\omega(\theta)$
is defined as the probability of finding an object in a solid angular element $d\Omega$ at 
an angular distance $\theta$ from another given object:
\begin{equation}
dP=n[1+\omega(\theta)]d\Omega
\end{equation}
where $n$ is the mean surface density of the objects in the sample. We calculate the auto-correlation 
function by the natural estimator:
\begin{equation}
1+\omega(\theta)=\frac{DD(\theta)}{RR(\theta)}
\end{equation} 
where $DD$ and $RR$ are the numbers of data-data and random-random pairs, respectively, at an 
angular distance $\theta$. The cross-correlation function is calculated by the simple estimator 
proposed by Peebles (1980):
\begin{equation}
1+\omega(\theta)=\frac{D_1D_2(\theta)}{D_1R_2(\theta)}
\end{equation}
where $D_1D_2$ is the number of cluster-GRB(z) pairs at an angular distance $\theta$, 
and $D_1R_1$ the number of cluster-random pairs at the same angular distance.  

Generally, a positive two-point correlation function means some association, and 
a negative function some avoidance. If the objects are uniformly distributed, 
the function will be zero. 
The calculated auto-correlation functions are shown in the left column in Figure 4, and 
the cross-correlation function in the right column. A Monte-Carlo simulation with 500 random 
realizations is performed to calculate each two-point correlation function and corresponding 
significance level. The random catalogs are generalized as the same method used in NNA. 
All the functions are obtained by binning the data into an angular bin size of 1\degr. 
As was the case previously reported in the NNA, no 
correlation between the clusters and the GRB(z) sample can be significantly 
identified in our current study. Significant result can be only identified for the
auto-correlation function of the clusters (with significance level $>90\%$). 
The clusters are clustered at small angular distance scales $\theta<3\degr$.

\begin{figure}[h!!!]
\centering
\includegraphics[width=9.0cm, angle=0]{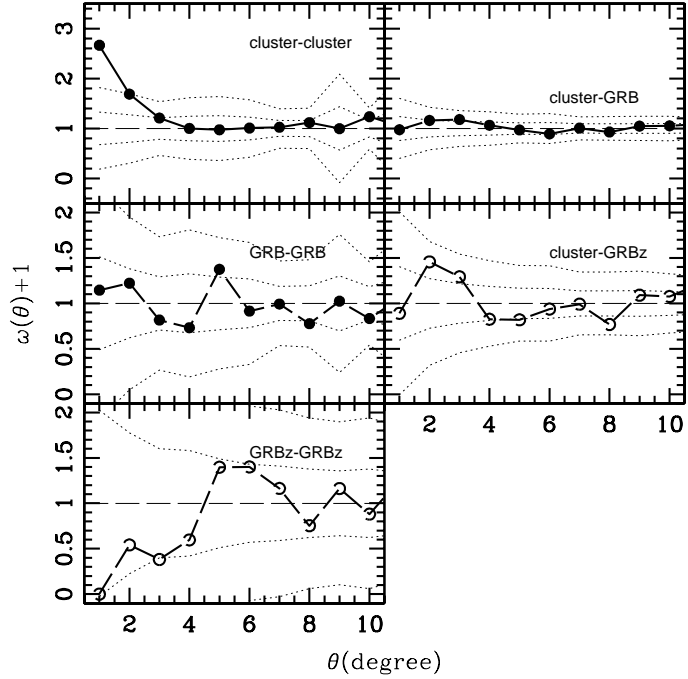}
\begin{minipage}[]{85mm}
\caption{
The left and right columns present the auto-correlation functions and cross-correlation functions, respectively.
All the functions are binned into a 
In each panel, the calculated functions are plotted by 
of the nearest-neighbor distance is presented by the solid line, and the Monte-Carlo simulation
with 500 random realizations by the dashed line. The two sets of thin dashed lines in each panel mark the
1$\sigma$ and 2.5$\sigma$ confidence limit derived from the simulation.
}
\end{minipage}
\label{Fig1}
\end{figure}

%


\section{Conclusion And Implication}
\label{sect:con}

With much improved burst locations down to a few arcseconds, the \it Swift\rm/XRT GRB sample
allows us to re-examine the correlation between GRBs and foreground galaxy clusters.
By using the NNA and angular two-point cross-correlation function,
our analysis indicates that there is no enough evidence supporting the correlation between the 
foreground X-ray selected \it ROSAT \rm brightest clusters and \it Swift\rm/XRT GRBs.

Although our study is not the first one examining the relation between GRBs and foreground galaxy clusters,
it is indeed the first study using the GRBs position information with accuracy down to a few arcseconds.
The contradiction of the results obtained before 2004 (see more details and the
references listed in Section 1) is likely due to the poor localization
of the BATSE GRB sample. It is note that the error boxes of the BATSE GRBs range from a fraction of a degree to as
large as $\sim30\degr$. 

Generally speaking, the motivation of searching the correlation between distant GRBs and nearby massive 
structure is to study the local large structure in terms of the weak lensing effect of the GRBs. 
Assuming the weak lensing effect indeed occurs for the GRBs, how could we understand the 
non-correlation revealed in the current study? In fact, the association due to weak lensing effect 
not only depends on the mass of the lenses, but also depends on the number-flux relation of 
the background objects. Weak lensing increases the brightness of faint objects, while expands
the area behind the lenses. Combining the competing two factors, the correlation function could be 
related with the number-flux relation as (e.g., Williams \& Irwin 1998; Myers et al. 2003) 
\begin{equation}
\omega(\theta)=\mu^{2.5\beta-1}-1
\end{equation} 
where the number-flux relation is approximately described as a powerlaw: $N(>\log f)\propto 10^{-2.5\beta\log f}$ 
(Boyle et al. 1988), $\mu$ is the magnification factor. 
This equation indicates that $\omega(\theta)=0$ when $\beta=0.4$.

The value of $\beta$ could be constrained from the \it Swift \rm/XRT observations. To estimate $\beta$, we 
use the sample recently compiled by Zheng et al. (2009). Zheng et al. (2009) collected the information of 
amount of GRBs observed by \it Swift \rm\ satellite before 2008 from the literature and on-line databases. 
The number-flux relation of these GRBs is displayed in Figure 5 for the 3\,keV X-ray flux density at 11 hour.    
The data are evenly binned into 9 bins in the logarithmic space.
The over-plotted error bars correspond to the 1$\sigma$ Poisson noise in each bin.
The solid line plots the best fitted smoothed powerlaw model with an expression:
\begin{equation}
\frac{dN}{d\log f}=\frac{N_0}{10^{a(\log f-\log f_0)}+10^{b(\log f-\log f_0)}}
\end{equation}  
where $N_0=218$ is the total number of the GRBs. Comparing the equation with Eq (5) yields 
$\beta=0.4b$ at the bright end. A weighted least-squared fitting yields the following 
parameters: $a=-0.0003\pm0.003$, $b=1.05\pm0.03$, and $\log f_0=1.55\pm0.02$. Although the sample 
seems to be clearly incomplete at the faint end, $\beta$ is inferred to be $0.42\pm0.01$ from the well sampled 
bright end. In fact, about half of the GRBs listed in the GRB sample are bright in X-ray with 
$\log(f_x/10^{-3} \mu\mathrm{Jy})>1.5$.
This exercise suggests that the zero correlation function of the GRBs is likely due to 
their number-flux relation, although the weak lensing effect may indeed occur for these GRBs.

\begin{figure}[h!!!]
\centering
\includegraphics[width=9.0cm, angle=0]{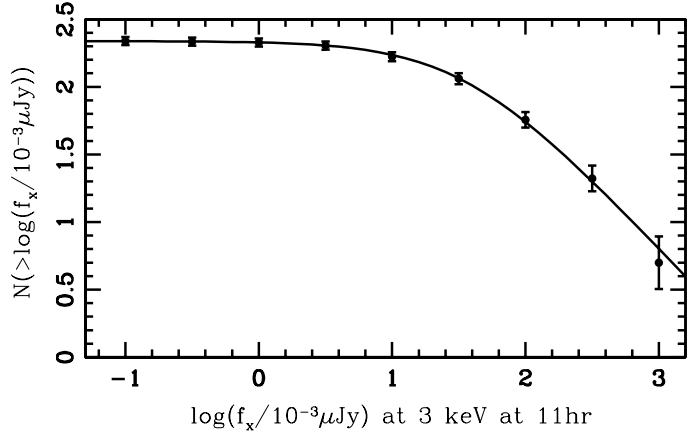}
\begin{minipage}[]{85mm}
\caption{
The number-flux relation for the GRBs compiled by Zheng et al. (2009). The data are 
evenly binned into 9 bins in the logarithmic space. The errorbars over-plotted are 1$\sigma$ Poisson noise.
The solid line is the best fitted smoothed powerlaw model (see Eq. 6). 
}
\end{minipage}
\label{Fig1}
\end{figure}



\begin{acknowledgements}
We are grateful to Professor Z. G. Deng for discussions and suggestions about the statistics
used in this paper. The authors thank the anonymous referee for comments and suggestions that 
improve the paper. 
This work was funded by the National Science Foundation of China under grant 10803008 and
National Basic Research Program of China (grant 2009CB824800), and supported by Chinese 
Academy of Science (under grant KJCXZ-YW-T19).
\end{acknowledgements}


\label{lastpage}

\end{document}